\documentclass[usenatbib]{nature}

\usepackage{amssymb}
\usepackage{url}
\usepackage{graphicx}

\newcommand\teff{T$_{\rm eff}$}
\newcommand{\msun}{\ensuremath{\, {M}_\odot}}

\newcommand{\wcen}{$\omega$\,Cen}

\newcommand{\Mc}{M$_{\rm c}$}
\newcommand{\Menv}{M$_{\rm env}$}

\newcommand{\Mturb}{M$_{\rm turb}$}
\newcommand{\Mrotev}{M$_{\rm ev}(\omega)$}
\newcommand{\Mcw}{M$_{\rm c}( \omega)$}
\newcommand{\deltaMrot}{$\Delta$M$_{\rm RGB}(\omega)$}

\newcommand{\xenvin}{X$_{\rm env-in}$}

\newcommand\aj{{Astron.~J}}%
\newcommand\apj{{Astrophys.~J}}%
\newcommand\apjl{{Astrophys.~J.~Letters}}%
\newcommand\apss{{Astrophys.~Space~Sci.}}%
\newcommand\aap{{Astron.~Astrophys.}}%
\newcommand\aapr{{Astron.~Astrophys.~Rev.}}%
\newcommand\mnras{{Mon.~Not.~R.~Astron.~Soc.}}%
\usepackage[usenames,dvipsnames]{color}

\title{Rapidly rotating second-generation progenitors for the blue hook stars of  $\omega$~Cen
}

\author{
Marco Tailo$^{1,2}$, Francesca D'Antona$^{1}$,  Enrico Vesperini$^{3}$,  Marcella Di Criscienzo$^{1}$, Paolo Ventura$^{1}$, Antonino P. Milone$^{4}$, Andrea Bellini$^{5}$, Aaron Dotter$^{4}$, Thibaut Decressin$^{1}$, Annibale D'Ercole$^{6}$, Vittoria Caloi$^{7}$, Roberto Capuzzo-Dolcetta$^{2}$
}

\begin{document}

\maketitle


\begin{affiliations}
 \item INAF- Osservatorio Astronomico di Roma, I-00040 Monte Porzio (Roma), Italy.
 \item Dipartimento di Fisica, Universit\'a degli Studi di Roma, La Sapienza, Roma, Italy.
 \item Department of Astronomy, Indiana University, Bloomington, IN (USA)
 \item Research School of Astronomy \& Astrophysics, Australian National University, Canberra ACT 2611, Australia
 \item Space Telescope Science Institute, 3700 San Martin Dr., Baltimore, MD 21218, USA
  \item INAF- Osservatorio Astronomico di Bologna, via Ranzani 1, I-40127 Bologna, Italy
 \item INAF, IAPS, Roma, via Fosso del Cavaliere 100, I-00133 Roma, Italy
\end{affiliations}

\begin{abstract}

Horizontal Branch stars belong to an advanced stage in the evolution of the oldest stellar galactic population, occurring either as field halo stars or grouped in globular clusters. 
The discovery of multiple populations in these clusters\cite{gratton2012, piotto2012}, that were previously believed to have single populations gave rise to the currently accepted theory
that the hottest horizontal branch members (the ‘blue hook’ stars, which had late helium-core flash ignition\cite{dcruz1996}, followed by deep mixing\cite{moehler2000,brown2001}) are the progeny of a helium-rich ``second generation" of stars\cite{dantona2002, brown2012}.
It is not known why such a supposedly rare event\cite{miller2008, brown2010}(a late flash followed by mixing) is so common that the blue hook of \wcen\ contains $\sim$30\% of horizontal branch stars\cite{andersonvdm2010}, or why the blue hook luminosity range in this massive cluster cannot be reproduced by models. 
Here we report that the presence of helium core masses up to $\sim$0.04 solar masses larger than the core mass resulting from evolution is required to solve the luminosity range problem. We model this by taking into account the dispersion in rotation rates achieved by the progenitors, whose pre-main sequence accretion disc suffered an early disruption in the dense environment of the cluster's central regions where second-generation stars form\cite{dercole2008}. Rotation may also account for frequent late-flash-mixing events in massive globular clusters.

\end{abstract}

In the colour–magnitude diagrams of globular clusters, the horizontal branch is the locus of core-helium-burning structures that are the progeny of red giant stars. For each such structure, the helium flash ignition has occurred either at the tip of the red giant branch, or anywhere along the evolutionary path that moves the star into the white dwarf stage\cite{brown2001}. In general no mixing occurs, and the result is a ‘standard’ horizontal branch structure: the smaller its hydrogen-rich envelope is, the hotter is the location of effective temperature (T$_{eff}$) in the model in the Hertzsprung–Russell diagram (that is, along the horizontal branch), up to about 32,000 K.

A very late helium ignition, along the white dwarf cooling pathway, can cause the helium core to mix with the small hydrogen-rich envelope\cite{sweigart1997,brown2001,cassisi2003} (late-flash-mixing), resulting in a slightly smaller helium-burning core plus a helium-rich, or a helium-dominated, envelope. Such ``blue hook" (BHk) structures attain a lower luminosity, and a higher \teff\ than do stars in the ‘extreme’ hottest horizontal branch standard locus, owing to the smaller opacity of the helium-rich atmosphere; they have been found in the \wcen\cite{whitney1994} and in a few other massive clusters\cite{brown2010}. \\

The blue hook in \wcen\ is particularly strikingis particularly striking because it contains about 30\% of the horizontal branch stars\cite{andersonvdm2010}, and it is extremely well defined in the Hubble Space Telescope (HST) observations. The optical colour-magnitude diagram\cite{andersonvdm2010} displays a blue hook that extends approximately 1.0 magnitudes in the F625W band, with a redder side much less populated (Extended Data Fig.~1a). A smaller sample of blue and UV data\cite{bellini2013} (Fig.~1a) displays a strong peak on the blue side of the colour distribution, where the helium-richer late-flash–mixed stars are located\cite{moehler 2007, latour2014}.\\
The models we adopt (see Methods) assume the helium core masses implied by the evolutionary process, and different initial envelope hydrogen abundances (\xenvin), together with an efficiency of helium settling calibrated to reproduce as well as possible the values of He/H versus \teff\ data from the literature\cite{latour2014} (Fig.~2). Tracks (that is, evolutionary sequences) with very low \xenvin\ values span only about 0.4 mag in F225W, but an extra 0.35 mag or so are obtained in models starting with larger \xenvin, as a result of the change in atmospheric composition from helium-dominated to hydrogen-rich (Extended Data Fig.~2 and Extended Data Table~1). 
Considering the whole range of \xenvin\ values explored, from 0.007 to 0.41, we can obtain a global extension of 0.9 mag in the F225W band (6th column of Extended Data Table~1), close to the range observed, but the tracks with \xenvin$> 0.06$\ do not match the colour-magnitude diagram well (Extended Data Fig.~4) nor the He/H versus \teff\ data (Fig.~2), and tend to merge with the extreme horizontal branch. For \xenvin$\leq$0.06, the tracks cover a maximum of  0.69~mag and the total extension is 0.79~mag.\\
Making the currently accepted assumption that the blue hook contains the progeny of the very-helium-rich second-generation stars, the existence of the blue hook stars can be attributed to late-flash–mixing events in models having an initial helium mass fraction of Y~=~0.37. Simulations including only objects having up to\xenvin=0.06 show the discrepancy with the observed luminosity range (Fig.~1b) \\
Modifications in model inputs have not much effect on the covered magnitude range:(1) standard flash-mixed models occur in a small core mass range  ($\delta {\rm M_{core}}<0.008$\msun)\cite{miller2008, brown2010}, corresponding to a negligible  magnitude difference in the tracks extension\cite{brown2010}; (2) the magnitude range can not be extended by increasing the adopted core overshooting parameters. 
3) we also have good reason to reject the possibility that the blue hook contains both the progeny of the standard and of the helium-rich populations (see discussion in Methods and Extended Data Fig.~3).
\\
The high-luminosity portion of the blue hook, at the correct colour location, can be covered by tracks having larger helium core masses M$_{c}$, as we show in Fig.~1a and Extended Data Fig.~1a for the track having $\delta$\Mc=+0.04\msun, and Y=0.37, where \msun\ is the solar mass. The helium flash ignites in more massive cores if very rapid core rotation delays the attainment of the flash temperatures. Of the (few) models available, we select a first-order approximation\cite{mengelgross1976} of models starting from solid-body rotation on the main sequence, and preserving angular-momentum in shells, to estimate the increase in the helium-flash core mass, $ \delta$\Mc, as a function of the initial angular velocity $\omega$. These models provide upper limits to $ \delta$\Mc ,because they do not allow for angular momentum transport and losses during main-sequence and post-main-sequence evolution. In this approximation, about half of the break-up main-sequence rotation rate (at which the centrifugal force at the equator becomes equal to the gravitational force, which is about $7 \times 10^{-4} {\rm s}^{-1}$) provides  a huge $\delta$\Mc=0.06\msun, leaving room for different \Mcw\ laws when available.\\
We propose that high rotation rates may be a consequence of the star-formation history and early dynamics in very massive clusters.
Specifically, in the model based on the formation of second-generation stars from the ejecta of asymptotic giant branch stars, a cooling flow
collects such ejecta in the central regions of the first-generation cluster and produces a centrally concentrated second-generation subcluster\cite{dercole2008, bekki2010}.
Observations showing that the helium-rich blue main-sequence stars—the progenitors of the blue hook stars—are spatially more concentrated than the rest of the main-sequence stars\cite{Bellini2009} and  retain some 'memory' of their initial spatial segregation,provide evidence that indeed such progenitors formed segregated in the innermost regions of the cluster, as predicted\cite{dercole2008}  and assumed in this study.\\

The contracting pre-main-sequence low-mass objects in the galactic disc (that is, stars like the variable star T Tauri)rotate with periods in the range from 2 days to 12 days.
The rotation rates of classical T Tauri stars do not increase with stellar age, because of magnetic disk-locking between the star and the disk\cite{armitageclarke1996}, which keeps the stellar rotation constant until the disk is lost. Stars in young clusters (such as $\alpha$Persei) show a wide distribution of rotation velocities, owing to differences in the time at which different objects break their magnetic coupling with the wind\cite{bouvier1997}. Rapidly rotating stars result from an early breaking of the of the coupling. Afterwards, during the main sequence, stellar winds slow down the rotation of the convective stellar envelopes, and old stars all appear to rotate slowly, but the inner core rotation is still fast despite the slow angular momentum transfer from the core to the envelope. We note that this model implies the presence of a crowded second generation.
\\
Figure~3a shows the fast decrease of the momentum of inertia of first- and second-generation stars in globular clusters (the latter ones have no initial deuterium, because the gas from which they formed was nuclearly processed at high temperature\cite{dercole2008}). Figure~3b describes the stellar angular velocity which can be obtained at the main sequence (age exceeding $2 \times 10^7$yr), as a function of the time at which the disk-star coupling is destroyed and evolution proceeds at constant angular momentum. The moment of inertia evolution of the case of a mass of M=0.7\msun, Y=0.35 is assumed,for rotation periods at detachment taken in the range 2-12 days. We see that there is ample model space to reach high angular velocities, provided that the second-generation (the progenitors of the blue hook stars) lose their disc at young ages, from 10$^5$yr (for the longer initial periods) to about $ 3\times 10^6$yr (for the short periods). 
\\
We model the dynamical encounters that destroy the disk, assuming that three encounters are necessary (see details in Methods), and use the $\delta$\Mc\ resulting from the timings of encounters as direct inputs for the simulation. The resulting distribution of the blue hook is shown in Fig.~1c and Extended Data Fig.~1b.
We remark that fast rotation will favour deep mixing in the giant envelopes during the last phases of evolution, enhancing the chemical anomalies of the most extreme second generation\cite{weiss2000,dantonaventura2007} and probably also favouring mixing at the onset of the flash: the late-flash-mixing event, although very different from what is described in one-dimensional models, will be more probable, thanks to the reduced entropy barrier between the core and the envelope, justifying the existence of the blue hook itself. In this scheme, the ‘non-mixed’ extreme horizontal branch stars should have been slowly
rotating, and in fact standard models match this group well. 
In the simulation we technically model them by assuming that the cool side of the hot horizontal branch is populated by the progeny of scarcely rotating (rotation rate $\omega < 10^{-6}$s$^{-1}$) stars (red squares in Fig. 1 and Extended Data Figs 1, 3 and 4). \Mc\ for this side of the simulation is the
mass of the non-rotating core, and in fact standard tracks match the extreme horizontal branch well.
 \\
The presence of rapidly rotating stars among the second-generation population should not be confined to modelling the blue hook in \wcen. Two conditions are at the basis of the success of the dynamical model used here: first, we must deal with second-generation stars, which form in a cooling flow at stellar densi-
ties much higher than that of the first-generation stars; and second, the population must be abundant, so it requires the presence of very massive clusters to start with. Of the most massive clusters, M54 (NGC~6715) shows a blue hook very similar in shape to that of \wcen\cite{piotto2014lp}, and NGC~2419 also has an extended blue hook\cite{dicrisci2015}. NGC~6388 and NGC~6441 have peculiarly extended horizontal branches, unlike other metal-rich clusters. It is certainly possible that their thick red horizontal branch (the ‘red clump’), modelled by assuming a large helium spread in the second generation\cite{caloi2007}, is also caused, in part, by the presence of larger core masses due to fast rotation. The reproduction of the horizontal branch morphology would then require, in the second generation, a smaller helium content increase than predicted by standard models\cite{caloi2007}, and this would be more consistent with the helium spread derived by the main-sequence colour thickness in NGC~6441\cite{bellini63882013}.  We note that a large rotation spread in this case does not produce a
prominent blue hook, but an anomalous red clump. NGC 2808, just a bit less massive, could be a borderline case, in which the presence of fast rotators is uncertain.\\
Smaller but important rotation ranges may be present in other less-massive clusters as well, and may have been due to the same process of early disk destruction in second-generation stars: in many clusters the mass loss necessary to explain the location of second-generation stars has to be slightly larger than in the first generation, possibly because of their rotation rate\cite{salaris2008, dantona2013}.\\
The connection between the magnitude extension of the blue hook stars in \wcen\ and the range of rotation rates of their progenitors might be another manifestation of the interplay between the formation of multiple populations in clusters, their dynamical behaviour and the properties of the component stars. At the same time, the model developed in this work for the blue hook supports the model for the dynamical and chemical formation of multiple populations\cite{dercole2008} based on the contribution of asymptotic giant branch stars and adopted in this study.
%

\vskip 30pt

\begin{addendum}
 \item[Acknowledgements] A.P.M. acknowledges support by the Australian Research Council through Discovery Early Career Researcher Award DE150101816. E.V. acknowledges support from grant NASANNX13AF45G. P.V. and F.D'A. acknowledge support from PRIN INAF 2011 '‘Multiple populations in globular clusters: their role in the Galaxy assembly''(principal investigator E. Carretta), and P.V. acknowledges support from PRIN MIUR 2010-2011, project ''The Chemical and Dynamical Evolution of the Milky Way and Local Group Galaxies'' (principal investigator F. Matteucci). T.D. acknowledges support from the UE Program (FP7/2007–2013) under grant agreement number 267251 of Astronomy Fellowships in Italy (ASTROFit). M.D.C. acknowledges support from INAF-OAR. A.B. acknowledges support from STScI grant AR-12656.

\item[Author Contributions] M.T., F.D'A., E.V. and M.D.C. jointly designed and coordinated this study. F.D'A. proposed and designed the rotational evolution model. E.V. designed and computed the dynamical simulation. M.T. and P.V. computed the new evolutionary models. A.D. computed synthetic colours with an internally consistent treatment of extinction across all bandpasses. M.T. and M.D.C. performed the simulations and the analysis. A.B. performed the data reduction and calibration for the WFC3/UVIS exposures. A.M. dealt with the optical data and the comparison with spectroscopic data.
T.D. dealt with the problems connected to the modelling of stellar rotation. V.C. contributed to the discussion and to the writing of the text. A.D’E. and R.C.-D. provided insight on the dynamical aspects. All authors read, commented on and approved submission of this article.

 \item[Competing Interests] The authors declare that they have no competing financial interests.
 \item[Correspondence] Correspondence and requests for materials
should be addressed to Francesca D'Antona~(email: francesca.dantona@oa-roma.inaf.it or franca.dantona@gmail.com).
\end{addendum}

\noindent {\bf FIGURES}

\begin{center}
\includegraphics[width=0.95\columnwidth]{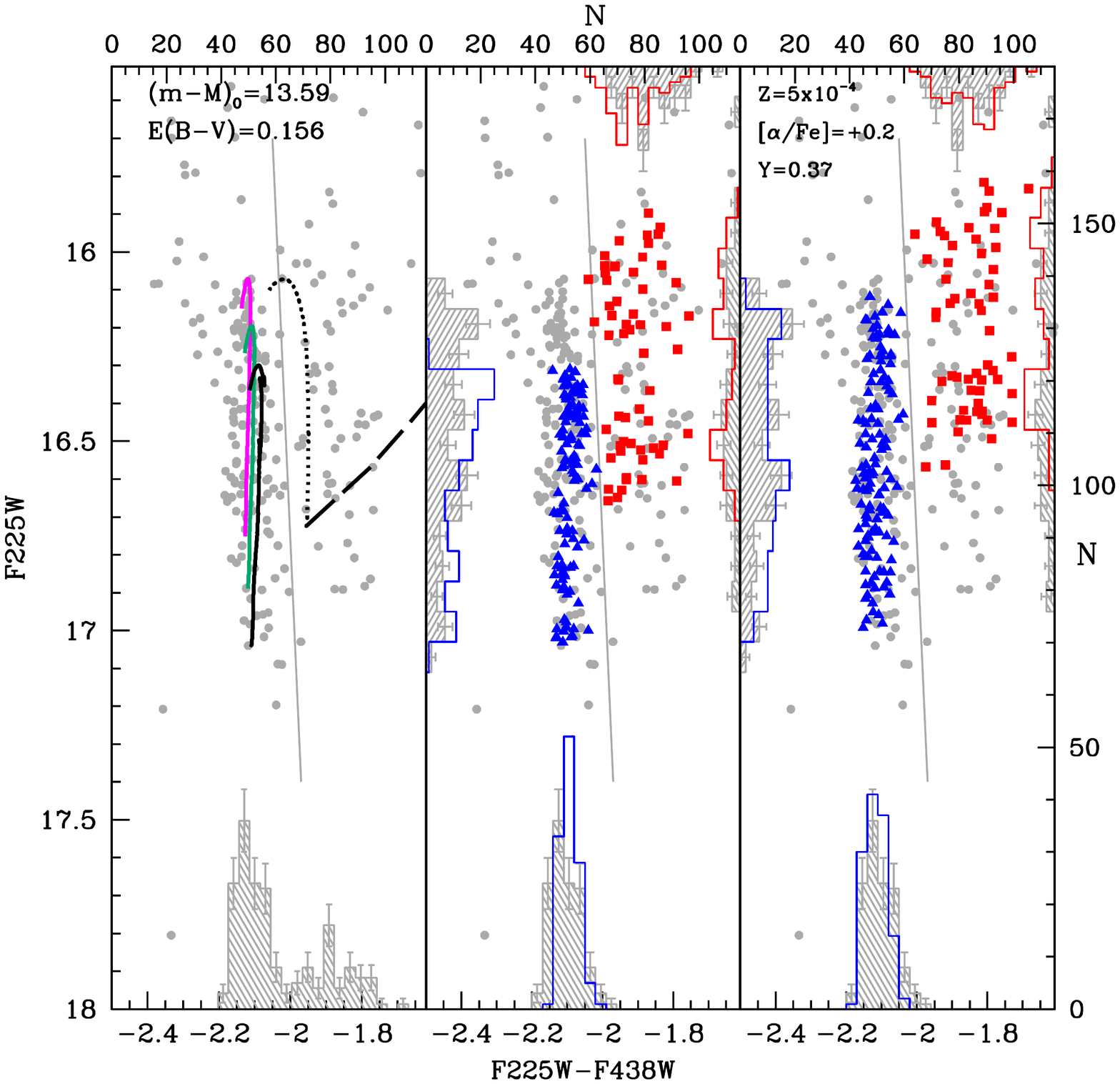}
\end{center}

\noindent {\bf Figure 1 - Simulations for the ultraviolet data.} \\
Grey dots are the observeds tars in \wcen.  Tracks plotted on the blue hook in \textbf{a} are: for $M=0.466\msun$ and Y=0.37 (black); for 0.489\msun, Y=0.25, standard \Mc (green);  and for \Mc\ increased by 0.04\msun (magenta). Models have  \xenvin=0.06. Also shown are the zero-age horizontal branch for Y=0.37 (dashed balck) and the 0.471\msun track (dotted black). Labels indicate bolometric distance modulus $(m-M)_0$ and reddening $E(B-V)$ adopted for the comparison. \textbf{b}, Simulation with a late-flash mixture of \xenvin stars for $Y = 0.37$ (see text for the details of the chemistry labels at the top). Comparison with data are shown in the histograms on the left (blue hook, blue triangles) and right (extreme horizontal branch, red squares) side. On the histograms, $1\sigma$ Poisson errors are shown with respect to the counts N. \textbf{c}, Simulation resulting from coupling dynamics and evolution (same mixture of \xenvin). Colour code and comparisons as in \textbf{b}.

\begin{center}
\includegraphics[width=0.95\columnwidth]{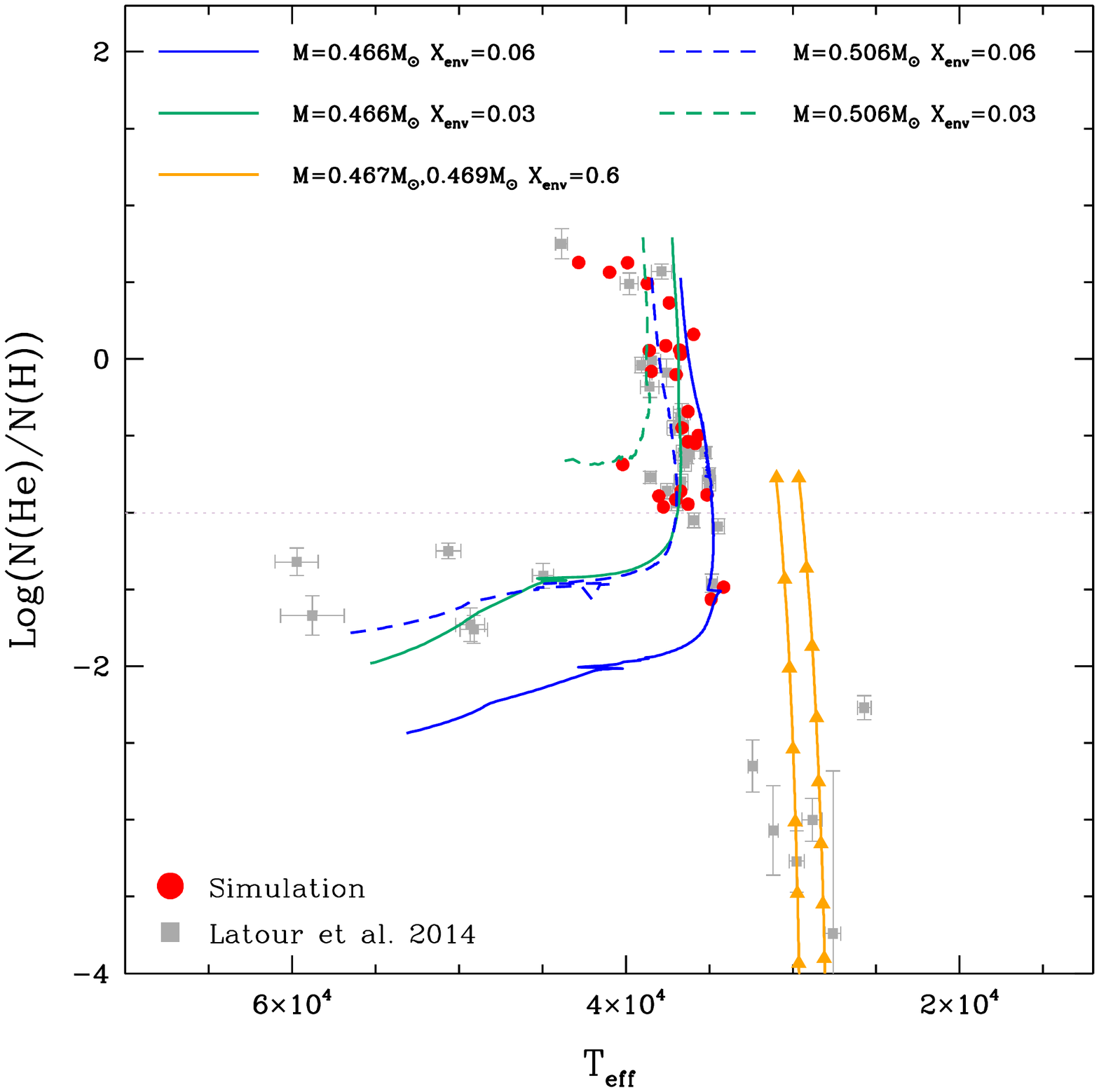}
\end{center}

\noindent {\bf Figure 2 - Calibration of H--He diffusion} \\
Grey squares show the He/H versus \teff data\cite{latour2014} with error bars. The evolution of three models is shown for different values of \xenvin . Diffusion is computed assuming that the envelope mass below which we assume diffusion is operating, \Mturb, is 10$^{-6}$\msun. The orange tracks show the evolution of the smallest non mixed tracks. Triangles mark each 10$^7$~yr interval, for \Mturb=10$^{-7}$. The red dots represent 26 stars randomly extracted from the simulation in Fig.~1c. Along the magenta curve, no star with \xenvin=0.21 are extracted, they would be cooler than the observed sample.

\begin{center}
\includegraphics[width=0.95\columnwidth]{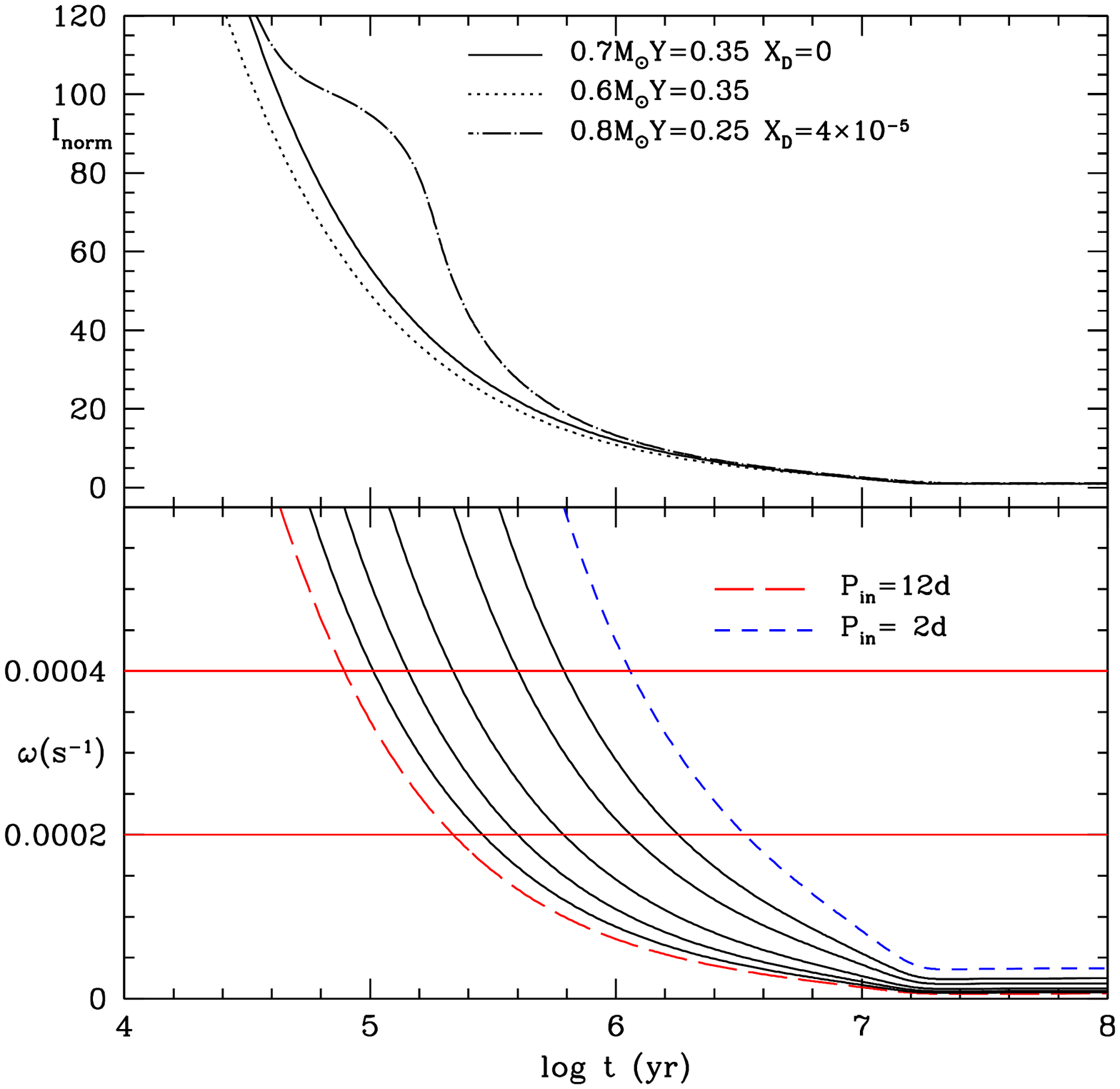}
\end{center}

\noindent {\bf Figure 3 - Rotational evolution from disc--detachment to the main sequence}\\
\textbf{a}, Time evolution of moment of inertia (I$_{\rm norm}$, in units of momentum at t=$2 \times 10^7$yr), for different masses, values of Y and  deuterium abundance X$_D$. \textbf{b}, For M=0.7\msun and Y=0.35, we show the rotation rate $\omega$\ at age t=2$\times 10^7$yr, as a function of the age at which the star detaches from the proto-stellar disk, and conservation of angular momentum begins. The lines represent different assumptions for the rotation period the disc is lost, from left to right: P$_{\rm in}$=12~days, 10~days, 8~days, 6~days, 4~days, 3 ~days and 2~days.  

\begin{methods}
\subsection{The data sets.}

We analyse optical\cite{andersonvdm2010} and ultraviolet\cite{bellini2013} Hubble Space Telescope data for the hot horizontal branch stars of \wcen.  Photometric errors are about 0.015~mag in filters F435W and F625W of ACS/WFC (Wide Field Channel of the Advanced Camera for Surveys), and 0.010~mag in the filters F225W and F438W of the WFC3. In the comparisons with optical data, we adopt distance moduli and reddening that fit the luminous part of the horizontal branch in these same data (M.T., PhD thesis, in preparation).  The values correspond to  $E(B-V)$=0.156 and $(m-M)_0$=13.63,  when using extinction ratios A$_{\rm F435W}$/A$_{\rm V}$=1.362 and A$_{\rm F625W}$/A$_{\rm V}$=0.868, explicitly computed for this work, and appropriate for stars of \teff$>$20000K, using standard extinction curves\cite{cardelli1989, odonnell1994}, and A$_{\rm V}$=3.1$\times E(B-V)$. The F225W and F438W data\cite{bellini2013}  have been recalibrated. Distance modulus and reddening are chosen by adjustment of the simulations results, and are labelled in Fig.~1. Using    A$_{\rm F225W}$/A$_{\rm V}$=2.670 and A$_{\rm F438W}$/A$_{\rm V}$=1.356, the bolometric distance modulus results to be $\sim$0.08~mag smaller than the modulus derived from optical data.\\
The prominent peak shown by number counts on the blue side (histogram at the bottom of Fig. 1a), followed by a sharp decline, and by a second broader peak at redder colours, corresponds to a change in the spectroscopic abundances:  the cool side of the distribution shows very helium-poor spectra\cite{moehler2007},  a feature ascribed to the effect of helium-diffusion in hydrogen-rich envelopes (the hottest standard horizontal branch stars). 
An almost abrupt transition to hydrogen-helium intermediate composition occurs at \teff$\simeq 35000$K\cite{moehler2007, latour2014}, where the sharp peak occurs. The contemporary presence of helium and carbon\cite{moehler2011, brown2012, latour2014} points out that the stars at \teff$\gtrsim$35000K are the result of late-flash-mixing\cite{brown2001}, so they must be blue hook stars. 
Standard one-dimensional simulations of late-flash-mixing end up with very small hydrogen abundance left in the envelope (X$\sim4 \times 10^{-4}$, see ref.  13). A ‘shallow’ mixing, leaving some percentage of hydrogen abundance, occurs only for a metal mass fraction that is ten times larger(Z$\geq$0.01 ref. 8) than the relevant metallicity in our sample. Nevertheless, the comparison with observed values of N(He)/N(H) requires that many blue hook stars are the result of a shallow mixing.  \\
We hand-draw a possible dividing line between blue hook and extreme, non mixed, horizontal branch stars. With this subdivision, in the optical sample the blue hook stars are $\sim$320 and the cooler stars are $\sim$150. In the UV sample we get 130 blue hook stars plus 60 cooler stars. \\

\subsection{Standard models.} 
Horizontal branch models are computed with the ATON code\cite{ventura2008code}.  Some useful results are given in Extended Data Tables 1 and 2. \\
Most models start from zero age horizontal branch, where the helium core mass is fixed by  previous evolution up to the helium flash, and the rest of the mass is in the hydrogen rich envelope. 
We also use as guidelines the results of some full late-flash evolutions, computed following the standard methodology\cite{castcast1993, dcruz1996} of  evolving the mass corresponding to an age of 12 billion years (Gyr) along the red giant branch, for increasing wind mass loss rates. Our results substantially confirm the previous findings\cite{cassisi2003, brown2001,miller2008}. The mass range of late flashers can be estimated in $\sim$0.02--0.03\msun, and our range of late flash mixed models is $\lesssim$0.005--0.006\msun\ (the largest mass range found in the literature is $\lesssim$0.008\msun ; ref.8). \\

\subsection{Models including sedimentation}
Guided by the evolutionary results, we  built up sets of models, characterized by three main parameters: core mass \Mc, envelope mass \Menv, and initial hydrogen mass abundance in the envelope \xenvin, and followed their evolution, starting from the zero age horizontal branch. 
As there are both helium-dominated and hydrogen-richer spectra among the BHk stars\cite{moehler2011, latour2014}, models need to include helium diffusion, to correctly derive the ultraviolet magnitudes, which are strongly dependent on the \teff\ of the models. An increase in the flux is expected, due to the shift in \teff\ that the star suffers when diffusion changes the helium carbon dominated atmosphere into a hydrogen-rich one.  The speed of diffusion, a byproduct of the  residual turbulence in the outer envelope\cite{michaud2007, michaud2008},  must be calibrated to be compatible with the observed values of hydrogen abundance in the blue hook stars\cite{latour2014,moehler2011}. In the models we calibrate the parameter \Mturb. Mass loss can play a concomitant role, and, when included,  \Mturb\ must be larger to remain consistent with the observations.  We show our choice among computations in  Extended Data Fig.~2. The theoretical (Extended Data Fig.~2b) and observational (Extended Data Fig.~2a) Hertzsprung-Russell diagram evolution of the tracks having \xenvin\ from 0.007 to 0.41 are plotted for two settings of diffusion efficiency. The F225W extension of the tracks increases from about 0.4~mag up to cover almost the full extension of the blue hook(histogram in Extended Data Fig. 2a), by increasing \xenvin up to the maximum value. 
Extended Data Fig. 2c shows the time evolution of \xenvin\ for the two choices of \Mturb. A better representation of spectroscopic observations requires use of the models with slower diffusion. Figure 2 shows the comparison of tracks and the location of simulated points (red dots) for a mixture of stars with different initial abundances, \xenvin = 0.007 (10\%), 0.03 (45\%) and 0.06 (45\%).  The \teff\ location and the He/H observed abundances are instead less compatible if we include a percentage of tracks having \xenvin$>$0.06 (see also the simulation comparisons).

\subsection{Assumptions for the rotating models.} 
In the blue hook progeny of rotating stars the core mass at the He--flash, \Mcw, is larger than the core mass of late-flash-mixing models computed by standard stellar evolution.  From models of evolution starting from solid-body rotation on the main sequence and angular-momentum conservation in shells\cite{mengelgross1976}, we derive a parabolic expression for the helium-flash core-mass increase $ \delta(M_{\rm core})$ as a function of the angular velocity $\omega$\ (in units of per second):
\begin{equation}
 \delta(M_{\rm core}/M_\odot)  = 3.86 \times 10^5 \cdot \omega^2  
\end{equation}
A self-consistent approach needs computation of rotating models starting from the zero age main sequence (where we can safely assume a rotation rate simply acquired by angular momentum conservation) and including all possible mechanisms of transfer of angular momentum from the core to the envelope, plus the loss of momentum of the envelope due to magnetic wind. In the models available in the literature\cite{gallet2013, palacios2003}, the parameters necessary to model these mechanisms have been calibrated from the atmospheric abundance variations induced by the associated chemical mixing, but most of the sampled stars were slowly rotating from the beginning, given that the fraction of young main-sequence stars which are fast rotating is quite small\cite{spada2011}. The coexistence of a fast-rotating core and slow-rotating envelope will lead to strong differential rotation that can be responsible for strong chemical mixing along the red giant branch, compared to the mixing possible for initially slower-rotating stars, an outcome that may explain extreme abundance anomalies in some clusters\cite{charbonnel2005,dantonaventura2007}, but it is not known how much these events may break down the inner fast core rotation. So we adopt the simple \Mc($\omega$) relation\cite{mengelgross1976} which does not take into account the core--envelope interactions.\\

The most important parameter in our investigation is the main sequence lifetime of these fast rotating stars, which is longer than the lifetime of non-rotating stars of the same mass. If the age is fixed, the rotating evolving mass \Mrotev\ will be larger than the non rotating mass. If a fraction of the rotating stars evolves through the helium flash, otherwise there would be no blue hook, we have to assume that its mass satisfies the relation: \\
 \Mrotev $ \gtrsim $ \Mcw + \deltaMrot \\
where  \deltaMrot\ is the total mass lost.  The flash luminosity in our models increases by $\delta$L$_{\rm flash}/\delta{\rm M_c}\simeq 1.9 \times 10^4$\ (where L and \Mc in solar units).  Assuming Reimers' mass loss rate, we expect an extra mass loss of 0.010-0.015\msun\ for each extra 0.01\msun\ increase in the core mass. If for example  we require $\delta$\Mc=0.04\msun, corresponding to $\omega \simeq 3.5 \times 10^{-4}$s$^{-1}$, the evolving mass must be $\sim$0.08--0.10\msun\ larger than the non-rotating mass, if it has to ignite the helium-flash. If we assume a stronger dependence of mass loss on L than in Reimers', we need larger \Mrotev.  Such an increase is compatible with the scarce existing estimates\cite{mengelgross1976} but a large computational effort is needed to solve the problem. A spread in the evolving mass  \Mrotev\ is also useful to meet, at least in a range of $\omega$, the strict requirements of the late-flash-mixing conditions and let the stars populate the blue hook: those outside the correct range will evolve into the helium-core white-dwarf remnants whose existence has been recently established\cite{bellini2013}.\\

The core mass vs. luminosity relation shows that the tip of the RGB is extended by $\delta \log L/L_\odot \simeq 0.14$ (0.35 bolometric mag) for the core mass increase of $\sim$0.04\msun. If the model developed here is correct, the brightest giants should belong to the He--rich population, contrary to standard models (the core flash occurs at slightly smaller luminosities in the high Y, non rotating, models).

\subsection{Model transformations: from the theoretical to the observational plane}
We use bolometric corrections for Z=0.0005 and [$\alpha$/Fe]=0.2 ([Fe/H]=--1.74) available for atmosphere models with standard hydrogen abundance\cite{castellikurucz2004}. 
A correction is necessary, as these spectra do not represent accurately the peculiar atmospheres of late-flash-mixed stars\cite{brown2001}, with strongly enhanced helium and carbon abundances at the surface.  At a given \teff, we correct the helium dominated model magnitudes making them fainter by 0.057mag in the F225W band, an estimate based on the comparison of fluxes in helium-and-carbon-rich and in hydrogen-rich model atmospheres\cite{brown2010}. We keep this correction until the helium surface abundance remains above Y=0.9, then we switch to the direct table correlations. This choice contributes to stretch in magnitude extension the tracks with 0.03$\leq$ \xenvin $\leq 0.06$.

\subsection{Standard Simulations}
Synthetic models  for the simulations shown in Fig.~1b follow standard guidelines\cite{dc2008}.  For the cases employing standard tracks, we assume that the red giant mass M$_{RG}$ at a fixed age is a function of helium, at assumed metallicity. The mass on the horizontal branch is M$_{HB}$=M$_{RG}$(Y)-$\Delta$M, where $\Delta$M is the mass lost during the red giant phase. In our case, assuming Y=0.37 and 12 Gyr we derive M$_{\rm RG}$=0.658 \msun.
The observed ratio of stars between late flash mixed and ``normal" extreme horizontal branch stars is reproduced by assuming that $\Delta$M has a Gaussian dispersion $\sigma$=0.008 around an average value $\Delta$M=0.19\msun. With this choice, the resulting 60 extreme horizontal branch stars that did not suffer mixing are shown in Fig. 1b as red squares, and  the 130 blue triangles are the late flash mixed stars. When we extract a mass smaller than the late flash mixed masses, the star is assumed to evolve into the helium-core white-dwarf.  With the chosen parameters we find 110 helium-core white dwarfs. 
Comparison is shown by histograms of counts as a function of color and magnitude, given separately for extreme horizontal branch and blue hook stars. The used bins are 0.03mag and 0.08mag for colours and magnitudes respectively. The error bars are the result of individual count Poisson error ($\sqrt{N}$).  The observational histograms are grey, the simulated ones are blue and red for the colour (scale on the right) and magnitude (scale on the top) distributions. 
\\
This modelling shows the inherent difficulty in building the blue hook: the mass range of extreme horizontal branch models (defined as the models at \teff$>$20000K) is of a few hundredths of \msun\ (from 0.03 to 0.07\msun, larger for a larger Y), while late flash mixing covers a mass range of only $\lesssim$0.008\msun. The most recent modelling of horizontal branches of GCs shows that the mass loss spread of each component of multiple populations must be very small, $< 0.01$\msun.  If the blue hook standard models are truly describing the blue hook stars in globular clusters, the mass lost by the red giant progenitor must be finely tuned, so that it covers precisely this tiny mass range. If the blue hook is made up by more than one of the globula cluster populations (groups of stars with different helium and metal content) such a fine tuning must have been successfully met twice. This is a key issue to interpret the simulations.
\\
First we assume that  all the extreme horizontal branch stars have  Y=0.37. We use this Y value because it is consistent with the requirements of the asymptotic giant branch model for the formation of the extreme populations in clusters\cite{ventura2011}. We use different sets of flash mixed models, with \xenvin=0.007, 0.03, 0.06, 0.21 and 0.41. The  simulations with \xenvin$\leq 0.06$\ are unsatisfactory, as they do not cover the whole BHk extension (Fig.~1b). Including a fraction of stars with \xenvin=0.21, the magnitude range is marginally better covered, but the colour agreement is worse (Extended Data Fig.~4b). The magnitude range is not extended significantly, because already the track having \xenvin=0.06 reaches high atmospheric hydrogen content along the evolution including hydrogen versus helium diffusion (Extended Data Fig.~2c).
Including even stars with \xenvin=0.41, these latter merge with the extreme horizontal branch  (Extended Data Fig. 4c). \\

Extended Data Fig.~3 shows the simulation according to which the BHk is made up by superposition of two different late flash mixing populations, with  Y=0.25 and  Y=0.37, respectively\cite{cassisi2009wcen}. $\Delta$M's \ and  $\sigma$\ values used are reported in the figure. {\sl The mass losses necessary to obtain masses in the late flash mixing range differ by $\sim$0.13\msun\  for the two different Y groups.} While it is already very difficult to accept that the mass-loss of a unique population is so well constrained that the blue hook is populated at all, this double constraint is an even more extreme assumption. In spite of this, the whole extension of the hook is not yet fully covered.\\

\subsection{Coupled dynamical and evolutionary simulations}

To estimate the distribution of rotation rates we use a simple semi-analytical model to calculate the distribution of times, $T_{\rm enc100}$, needed for second-generation star-disk systems to have a  close encounter with another star at a distance $d<100$ astronomical units (AU) (the estimated dimension of the protostellar disk) and assume one - or more - such collisions to be able to break the magnetic disc-locking (in this case $T_{\rm enc100}$  is the detachment time). In order to calculate  $T_{\rm enc100}$ we have followed the orbits of 50,000 particles distributed as a King model with a concentration $c\simeq 1.7 $ and integrated the collision rate (see Eq. 7.194 in ref. 46) along the orbit to estimate $T_{\rm enc100}$. The system we are modelling is meant to represent the dense second-generation subcluster.  For our calculation we assume its total mass to be equal to $10^{5.5} M_{\odot}$ and half-mass radius of about one parsec. 
From the distribution of $T_{\rm enc100}$ (Extended Data Fig.~5a) we directly derive the rotation rates assuming the conservation of angular momentum from Fig.~3, and the corresponding increase in the core-mass from Eq.~1 in Methods (Extended Data Fig.~5c).  The initial periods P$_{\rm {in}}$ at detachment from the disk are randomly extracted in the range 3$<$ P(days)$<$12, using a  normal distribution centered at P=6~days, with standard deviation  $\sigma$=2~days. The distribution of the stellar rotation periods result of the simulation at detachment from the disc are given in Extended Data~Fig.~5b. The distribution is not Gaussian, as it also depends on the limitations we have imposed on the maximum $\delta {\rm M_c}$.
\\
For the photometric simulation, we then assume the dispersion in core-masses inferred by the  described simulation (Extended Data Fig.~5c). We are implicitly assuming that the increase in initial evolving mass due to rotation is enough to accommodate the increase both in core mass and mass loss in such a way that the stars are able to arrive at late-flash-mixing ignition for rotation rates $\omega >10^{-6}$s$^{-1}$. We also impose that a tail of slower rotating stars populate the extreme horizontal branch of non--flash mixed Y=0.37 models. 

\subsection{Code availability}

For the stellar evolution calculations we used ATON (not publicly available). For the N-body simulations we used Starlab (publicly available at http://www.sns.ias.edu/,starlab/). For the synthetic populations, we used a program written explicitly for this work (not publicly available).

\end{methods}

\noindent
{\bf ED Tables}

\begin{center}

\begin{tabular}{c c c c c c c c}     
\hline
\hline  
X$_{\rm env-in}$ & $F225W$ & $F225W$ & F225W & $\delta$F225W & $\delta$F225W  & $(F225W-F438W)_0$ &  $T_{eff}$ \\
                 & ZAHB & Peak & TAHB & (track) & total  & ZAHB &  ZAHB \\

\hline
 $0.007$ & +2.29 & +1.86 & +1.97&  0.43 & 0.43 & -2.734 & 37660  \\
 $0.030$ & +2.21 & +1.61 & +1.69&  0.60 & 0.68 & -2.731 & 37200  \\
 $0.060$ & +2.19 & +1.50 & +1.56&  0.69 & 0.79 & -2.726 & 36700  \\
 $0.210$ & +2.10 & +1.43 & +1.46&  0.67 & 0.83 & -2.693 & 34500  \\
 $0.410$ & +2.03 & +1.38 & +1.39&  0.65 & 0.90 & -2.603 & 31600  \\
\hline
\multicolumn{8}{c}{$\Delta M_{core}=+0.04 M_{\odot}$}\\
\hline
 $0.007$ & +1.96 & +1.58 & +1.76 &  0.38 & 0.71 & -2.700 & 39400  \\
 $0.030$ & +1.93 & +1.38 & +1.71 &  0.55 & 0.91 & -2.697 & 38900  \\
 $0.060$ & +1.91 & +1.27 & +1.35 &  0.64 & 1.02 & -2.693 & 38400  \\
\hline
\end{tabular}

\end{center}

\noindent
Extended Data Table 1~~-~~ {\bf Range in magnitude F225W for M=0.466\msun, M$_{\rm core}$=0.463\msun}\\
For each initial envelope starting H-abundance (col. 1), we give F225W band values of late flash mixed models at the zero age horizontal branch (col. 2), at the maximum luminosity (col. 3) and at the terminal point (col. 4) (defined as the model in which the core helium abundance becomes zero), and the magnitude interval spanned by each model  (col. 5) and by a mixture (col. 6). The mass of the model is 0.466~M$_{\odot}$, with an envelope mass of 0.003~M$_{\odot}$, thus the model core mass is $0.463 M_{\odot}$.

\begin{center}
\begin{tabular}{c c c c c c c c}     
\hline
\hline  
$Y$ & M$_{\rm Tip}$& $M_{HB_{min}}$ &  M$_{\rm core}$ & $M_{LFM}$ & M$_{\rm core}LFM$ & $\Delta_{MLF}$\\
\hline
0.25 & 0.812 & 0.500 & 0.496 & 0.489 & 0.486 & 0.323\\
0.37 & 0.657 & 0.471 & 0.469 & 0.466 & 0.463 & 0.191\\ 
0.40 & 0.620 & 0.466 & 0.465 & 0.462 & 0.459 & 0.158\\ 
\hline 
\hline
\end{tabular}
\end{center}

\noindent
Extended Data  Table 2~~-~~{\bf Data at 12 Gyr for models of Z=0.0005, [$\alpha$/Fe]=0.2}\\
We list  some important quantities from our isochrones of 12~Gyr, Metallicity Z=0.0005 and [$\alpha$/Fe]=0.2. Col. 1: helium mass fraction; col.2: evolving mass at the tip of the red giant branch (no mass loss); col.3: minimum ``standard" horizontal branch mass; col. 4: core mass in \msun, when the He--flash is ignited at the red giant branch tip; col. 5: mass for which we have late flash mixing; col. 6: core mass for the late flash mixed model; col 7: mass loss needed to achieve late flash mixing.

\noindent
{\bf ED  Figures}

\begin{center}
\includegraphics[width=0.95\columnwidth]{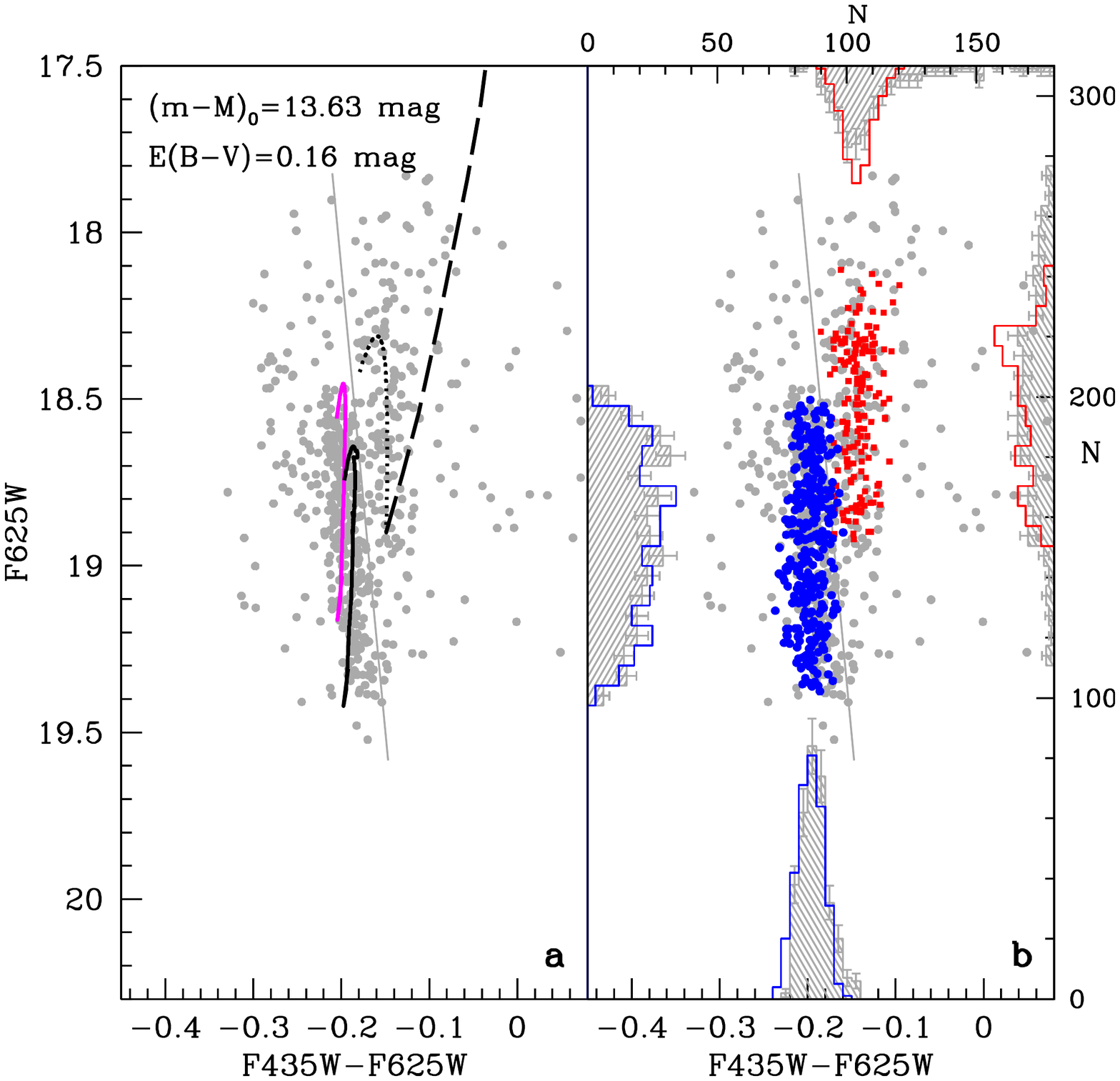}
\end{center}

\noindent
Extended Data  Figure~1 ~~{\bf Comparison simulation vs. optical data}\\
The observed data\cite{andersonvdm2010} are shown as grey dots. The diagonal line suggests the division between blue hook and cooler extreme horizontal branch stars. \textbf{a}: Tracks shown are for M=0.466\msun,  Y=0.37 and evolutionary \Mc=0.463\msun (black solid line), plus the corresponding track in which \Mc\ is increased by 0.04\msun \ (magenta solid line). \textbf{b}: Result of the coupled dynamical--evolutionary simulation (detailed explanation in Fig.~1).

\begin{center}
\includegraphics[width=0.95\columnwidth]{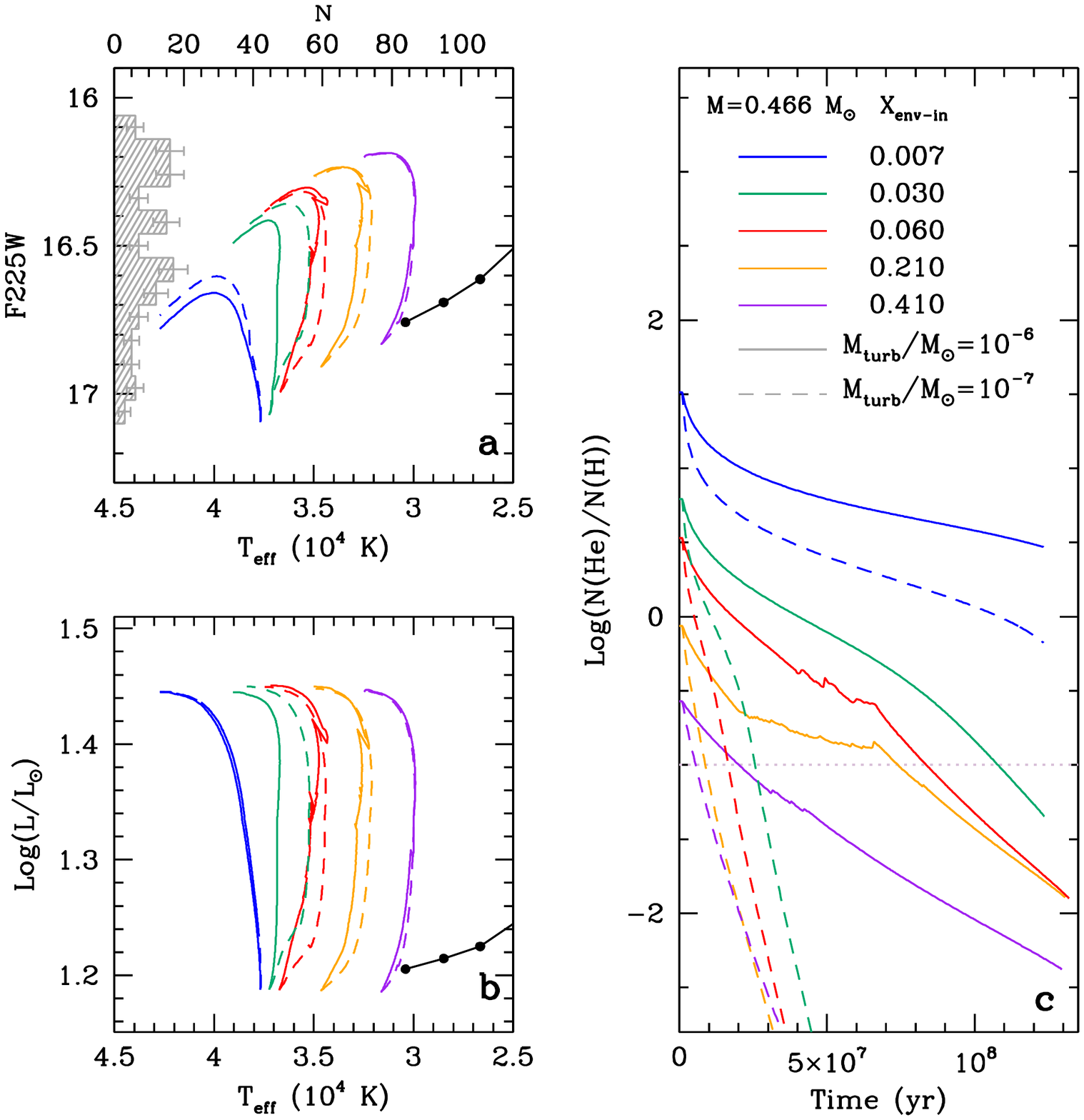}
\end{center}

\noindent
Extended Data Figure~2 ~~{\bf Models with He vs. H diffusion}\\
\textbf{a} and \textbf{b}  show a comparison of blue hook tracks from  \xenvin=0.007 to \xenvin=0.41, from left to right;  \Mturb=$10^{-6}$\msun\ (full lines) and $10^{-7}$\msun\ (dashed lines).  The bolometric luminosity (\textbf{b}) and  the F225W magnitude (\textbf{a}) are shown as function of \teff. The hot side of the zero age horizontal branch for Y=0.37 is shown (solid black line with dots). \textbf{c}: Evolution with time of the surface N(He)/N(H) due to diffusion. The solar ratio is shown as a horizontal grey dotted line. The observations of blue hook stars show that the slower diffusion is a better description (Fig.~2).

\begin{center}
\includegraphics[width=0.95\columnwidth]{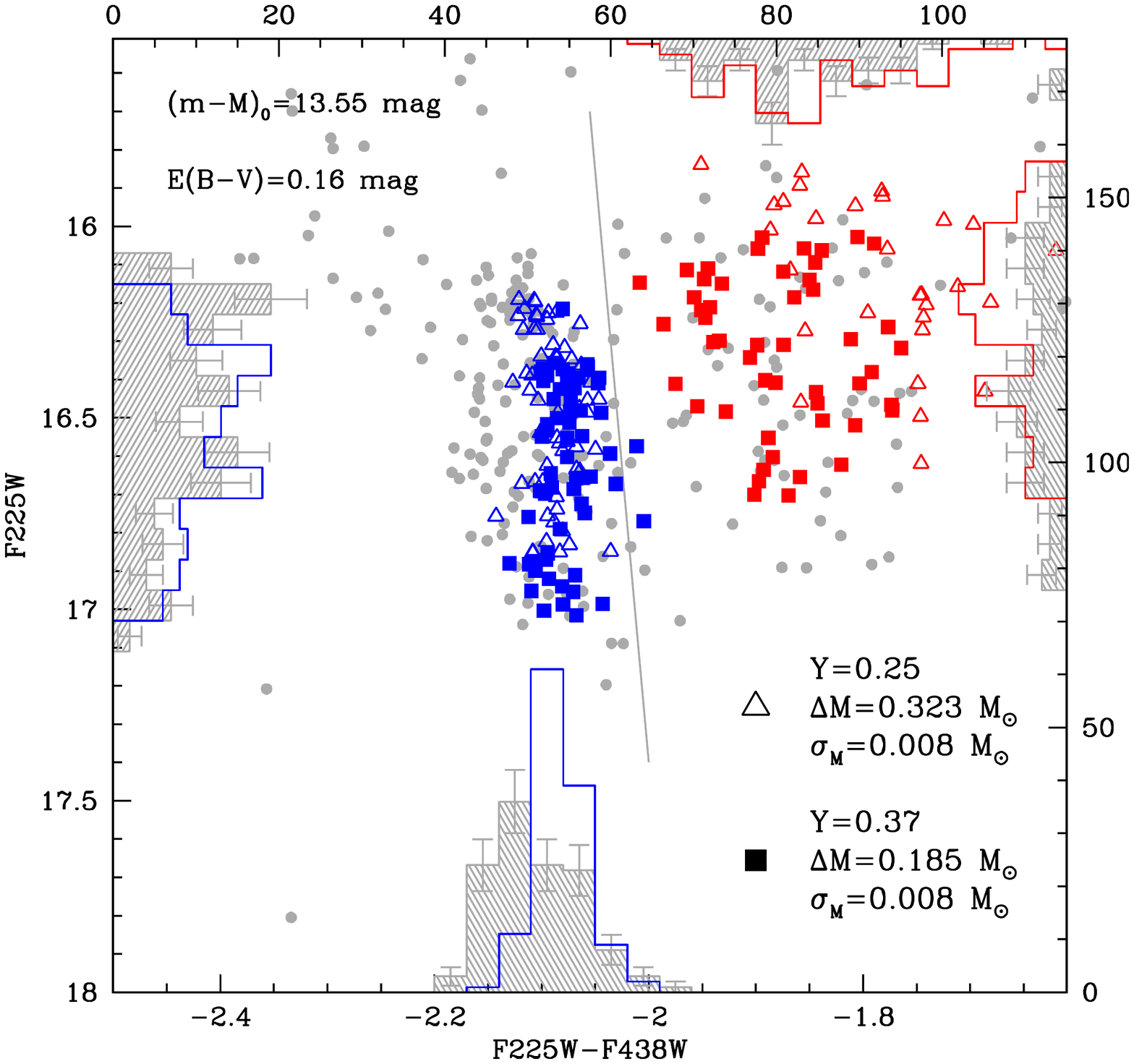}
\end{center}

\noindent
Extended Data Figure~3 ~~{\bf Simulation vs. data, including first and second generation stars}\\ 
As in Fig.~1 and Extended Data Fig.~1, but the ultraviolet data are compared with a simulation assuming that 50\% of blue hook stars are progeny of the Y=0.37 population (squares), and 50\% are progeny of the Y=0.25 population (triangles). The error bars are the result of individual one sigma Poisson error on the stellar counts.

\begin{center}
\includegraphics[width=0.95\columnwidth]{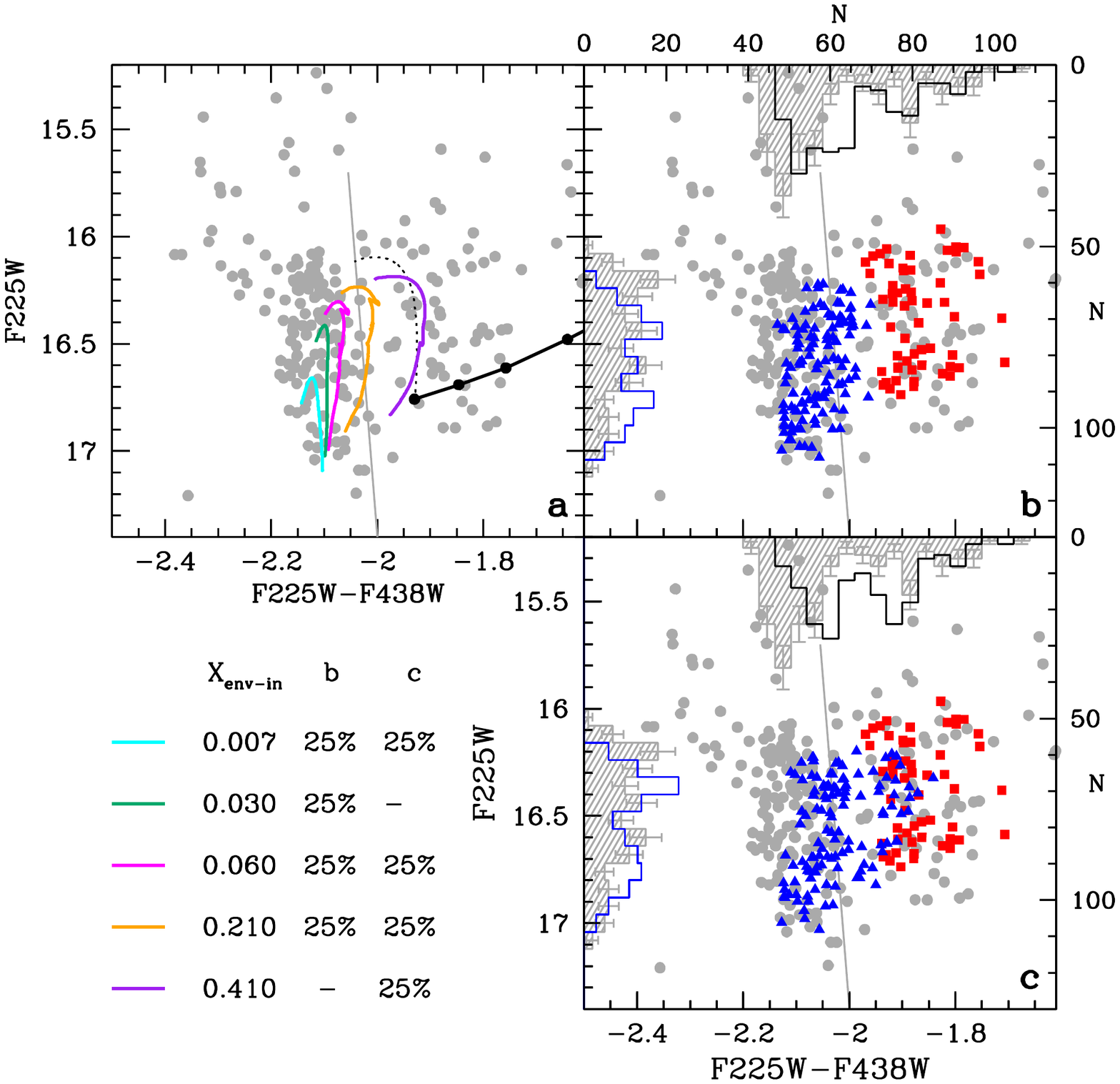}
\end{center}

\noindent
Extended Data Figure~4 ~~{\bf Tracks and simulation vs. data including models having \xenvin$>$0.06}\\ 
As in Fig.~1. \textbf{a}: tracks with different \xenvin\ and the end of the zero age horizontal branch with the location of the minimum non-mixed track for Y=0.37. \textbf{b}: simulation including  \xenvin\ up to 0.21, for the listed percentages of stars. \textbf{c}: simulation including also stars with \xenvin=0.41. In the latter simulation, the main aim is to show that the most hydrogen rich stars are located on the extreme horizontal branch, and not on the blue hook.

\begin{center}
\includegraphics[width=0.95\columnwidth]{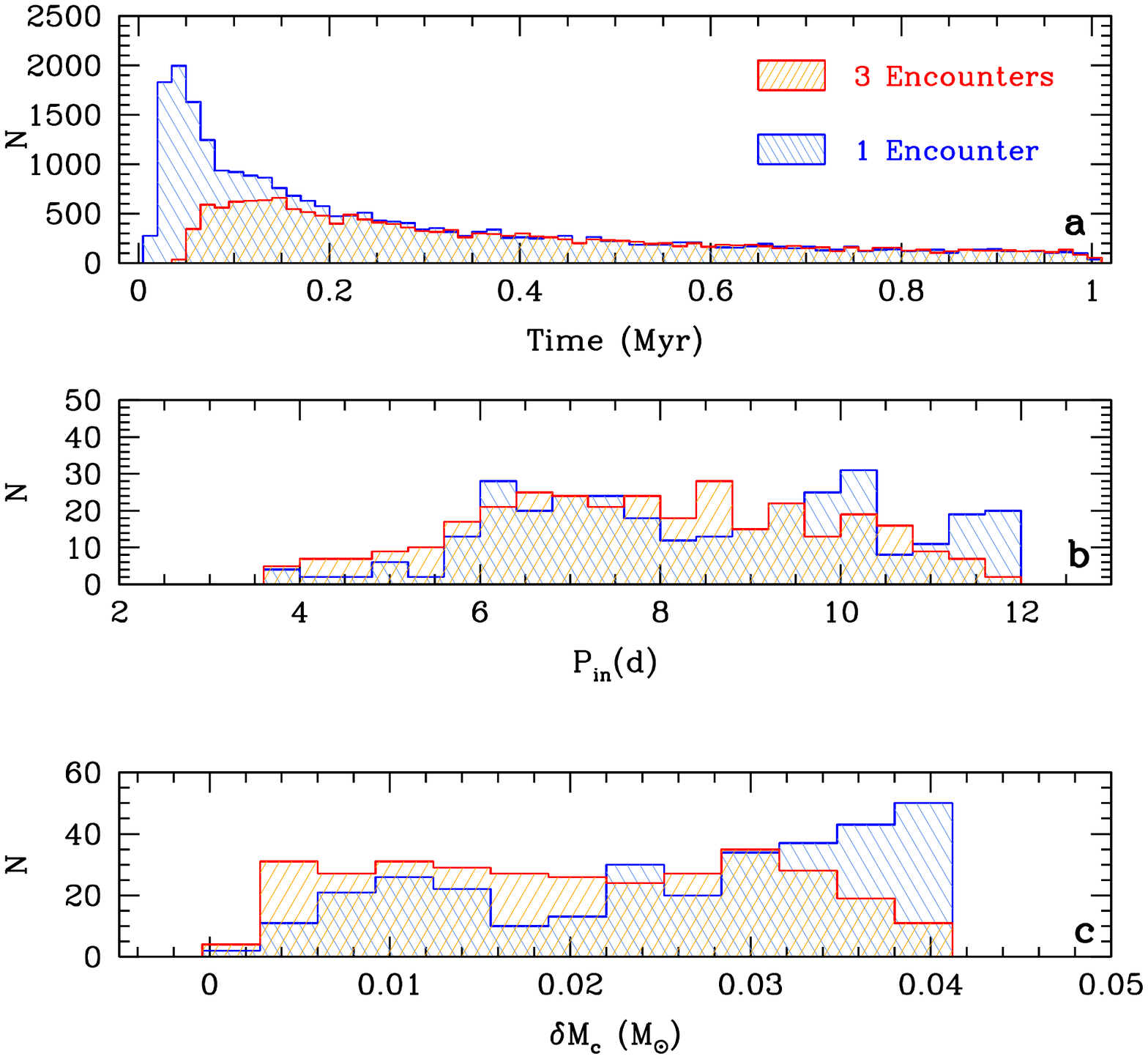}
\end{center}

\noindent
Extended Data Figure~5 ~~{\bf The dynamical model.} 
Two different cases of dynamical interactions are compared: either we make the hypothesis that the magnetic coupling of the disc is destroyed by one encounter with another star, at a distance entering the accretion disc ($<$100a.u), or that three of such encounters are necessary. \textbf{a}: histogram of encounters versus time of detachment. \textbf{b}: initial rotation period distribution employed in the two cases; \textbf{c}: distribution of the core mass increase which enters in the photometric simulation.


\vskip 30pt
\begin{thebibliography}{10}
\expandafter\ifx\csname url\endcsname\relax
  \def\url#1{\texttt{#1}}\fi
\expandafter\ifx\csname urlprefix\endcsname\relax\def\urlprefix{URL }\fi
\providecommand{\bibinfo}[2]{#2}
\providecommand{\eprint}[2][]{\url{#2}}

\bibitem{gratton2012}
\bibinfo{author}{{Gratton}, R.~G.}, \bibinfo{author}{{Carretta}, E.} \&
  \bibinfo{author}{{Bragaglia}, A.}
\newblock \bibinfo{title}{{Multiple populations in globular clusters. Lessons
  learned from the Milky Way globular clusters}}.
\newblock \emph{\bibinfo{journal}{\aapr}} \textbf{\bibinfo{volume}{20}},
  \bibinfo{pages}{50} (\bibinfo{year}{2012}).

\bibitem{piotto2012}
\bibinfo{author}{{Piotto}, G.} \emph{et~al.}
\newblock \bibinfo{title}{{Hubble Space Telescope Reveals Multiple Sub-giant
  Branch in Eight Globular Clusters}}.
\newblock \emph{\bibinfo{journal}{\apj}} \textbf{\bibinfo{volume}{760}},
  \bibinfo{pages}{39} (\bibinfo{year}{2012}).

\bibitem{dcruz1996}
\bibinfo{author}{{D'Cruz}, N.~L.}, \bibinfo{author}{{Dorman}, B.},
  \bibinfo{author}{{Rood}, R.~T.} \& \bibinfo{author}{{O'Connell}, R.~W.}
\newblock \bibinfo{title}{{The Origin of Extreme Horizontal Branch Stars}}.
\newblock \emph{\bibinfo{journal}{\apj}} \textbf{\bibinfo{volume}{466}},
  \bibinfo{pages}{359} (\bibinfo{year}{1996}).

\bibitem{moehler2000}
\bibinfo{author}{{Moehler}, S.}, \bibinfo{author}{{Sweigart}, A.~V.},
  \bibinfo{author}{{Landsman}, W.~B.} \& \bibinfo{author}{{Heber}, U.}
\newblock \bibinfo{title}{{Hot HB stars in globular clusters - Physical
  parameters and consequences for theory --- V. Radiative levitation versus
  helium mixing}}.
\newblock \emph{\bibinfo{journal}{\aap}} \textbf{\bibinfo{volume}{360}},
  \bibinfo{pages}{120--132} (\bibinfo{year}{2000}).

\bibitem{brown2001}
\bibinfo{author}{{Brown}, T.~M.}, \bibinfo{author}{{Sweigart}, A.~V.},
  \bibinfo{author}{{Lanz}, T.}, \bibinfo{author}{{Landsman}, W.~B.} \&
  \bibinfo{author}{{Hubeny}, I.}
\newblock \bibinfo{title}{{Flash Mixing on the White Dwarf Cooling Curve:
  Understanding Hot Horizontal Branch Anomalies in NGC 2808}}.
\newblock \emph{\bibinfo{journal}{\apj}} \textbf{\bibinfo{volume}{562}},
  \bibinfo{pages}{368--393} (\bibinfo{year}{2001}).

\bibitem{dantona2002}
\bibinfo{author}{{D'Antona}, F.}, \bibinfo{author}{{Caloi}, V.},
  \bibinfo{author}{{Montalb{\'a}n}, J.}, \bibinfo{author}{{Ventura}, P.} \&
  \bibinfo{author}{{Gratton}, R.}
\newblock \bibinfo{title}{{Helium variation due to self-pollution among
  Globular Cluster stars. Consequences on the horizontal branch morphology}}.
\newblock \emph{\bibinfo{journal}{\aap}} \textbf{\bibinfo{volume}{395}},
  \bibinfo{pages}{69--75} (\bibinfo{year}{2002}).

\bibitem{brown2012}
\bibinfo{author}{{Brown}, T.~M.} \emph{et~al.}
\newblock \bibinfo{title}{{Flash Mixing on the White Dwarf Cooling Curve:
  Spectroscopic Confirmation in NGC 2808}}.
\newblock \emph{\bibinfo{journal}{\apj}} \textbf{\bibinfo{volume}{748}},
  \bibinfo{pages}{85} (\bibinfo{year}{2012}).

\bibitem{miller2008}
\bibinfo{author}{{Miller Bertolami}, M.~M.}, \bibinfo{author}{{Althaus},
  L.~G.}, \bibinfo{author}{{Unglaub}, K.} \& \bibinfo{author}{{Weiss}, A.}
\newblock \bibinfo{title}{{Modeling He-rich subdwarfs through the hot-flasher
  scenario}}.
\newblock \emph{\bibinfo{journal}{\aap}} \textbf{\bibinfo{volume}{491}},
  \bibinfo{pages}{253--265} (\bibinfo{year}{2008}).

\bibitem{brown2010}
\bibinfo{author}{{Brown}, T.~M.} \emph{et~al.}
\newblock \bibinfo{title}{{The Blue Hook Populations of Massive Globular
  Clusters}}.
\newblock \emph{\bibinfo{journal}{\apj}} \textbf{\bibinfo{volume}{718}},
  \bibinfo{pages}{1332--1344} (\bibinfo{year}{2010}).

\bibitem{andersonvdm2010}
\bibinfo{author}{{Anderson}, J.} \& \bibinfo{author}{{van der Marel}, R.~P.}
\newblock \bibinfo{title}{{New Limits on an Intermediate-Mass Black Hole in
  Omega Centauri. I. Hubble Space Telescope Photometry and Proper Motions}}.
\newblock \emph{\bibinfo{journal}{\apj}} \textbf{\bibinfo{volume}{710}},
  \bibinfo{pages}{1032--1062} (\bibinfo{year}{2010}).

\bibitem{dercole2008}
\bibinfo{author}{{D'Ercole}, A.}, \bibinfo{author}{{Vesperini}, E.},
  \bibinfo{author}{{D'Antona}, F.}, \bibinfo{author}{{McMillan}, S.~L.~W.} \&
  \bibinfo{author}{{Recchi}, S.}
\newblock \bibinfo{title}{{Formation and dynamical evolution of multiple
  stellar generations in globular clusters}}.
\newblock \emph{\bibinfo{journal}{\mnras}} \textbf{\bibinfo{volume}{391}},
  \bibinfo{pages}{825--843} (\bibinfo{year}{2008}).

\bibitem{sweigart1997}
\bibinfo{author}{{Sweigart}, A.~V.}
\newblock \bibinfo{title}{{Helium Mixing in Globular Cluster Stars}}.
\newblock In \bibinfo{editor}{{Philip}, A.~G.~D.}, \bibinfo{editor}{{Liebert},
  J.}, \bibinfo{editor}{{Saffer}, R.} \& \bibinfo{editor}{{Hayes}, D.~S.}
  (eds.) \emph{\bibinfo{booktitle}{The Third Conference on Faint Blue Stars}},
  \bibinfo{pages}{3} (\bibinfo{year}{1997}).

\bibitem{cassisi2003}
\bibinfo{author}{{Cassisi}, S.}, \bibinfo{author}{{Schlattl}, H.},
  \bibinfo{author}{{Salaris}, M.} \& \bibinfo{author}{{Weiss}, A.}
\newblock \bibinfo{title}{{First Full Evolutionary Computation of the Helium
  Flash-induced Mixing in Population II Stars}}.
\newblock \emph{\bibinfo{journal}{\apjl}} \textbf{\bibinfo{volume}{582}},
  \bibinfo{pages}{L43--L46} (\bibinfo{year}{2003}).

\bibitem{whitney1994}
\bibinfo{author}{{Whitney}, J.~H.} \emph{et~al.}
\newblock \bibinfo{title}{{Far-ultraviolet photometry of the globular cluster
  omega CEN}}.
\newblock \emph{\bibinfo{journal}{\aj}} \textbf{\bibinfo{volume}{108}},
  \bibinfo{pages}{1350--1363} (\bibinfo{year}{1994}).

\bibitem{bellini2013}
\bibinfo{author}{{Bellini}, A.} \emph{et~al.}
\newblock \bibinfo{title}{{A Double White-dwarf Cooling Sequence in {$\omega$}
  Centauri}}.
\newblock \emph{\bibinfo{journal}{\apjl}} \textbf{\bibinfo{volume}{769}},
  \bibinfo{pages}{L32} (\bibinfo{year}{2013}).

\bibitem{moehler2007}
\bibinfo{author}{{Moehler}, S.} \emph{et~al.}
\newblock \bibinfo{title}{{The hottest horizontal-branch stars in {$\omega$}
  Centauri. Late hot flasher vs. helium enrichment}}.
\newblock \emph{\bibinfo{journal}{\aap}} \textbf{\bibinfo{volume}{475}},
  \bibinfo{pages}{L5--L8} (\bibinfo{year}{2007}).

\bibitem{latour2014}
\bibinfo{author}{{Latour}, M.} \emph{et~al.}
\newblock \bibinfo{title}{{A Helium-Carbon Correlation on the Extreme
  Horizontal Branch in {$\omega$} Centauri}}.
\newblock \emph{\bibinfo{journal}{\apj}} \textbf{\bibinfo{volume}{795}},
  \bibinfo{pages}{106} (\bibinfo{year}{2014}).

\bibitem{mengelgross1976}
\bibinfo{author}{{Mengel}, J.~G.} \& \bibinfo{author}{{Gross}, P.~G.}
\newblock \bibinfo{title}{{Possible effects of internal rotation in low-mass
  stars}}.
\newblock \emph{\bibinfo{journal}{\apss}} \textbf{\bibinfo{volume}{41}},
  \bibinfo{pages}{407--415} (\bibinfo{year}{1976}).

\bibitem{bekki2010}
\bibinfo{author}{{Bekki}, K.}
\newblock \bibinfo{title}{{Rotation and Multiple Stellar Population in Globular
  Clusters}}.
\newblock \emph{\bibinfo{journal}{\apjl}} \textbf{\bibinfo{volume}{724}},
  \bibinfo{pages}{L99--L103} (\bibinfo{year}{2010}).

\bibitem{Bellini2009}
\bibinfo{author}{{Bellini}, A.} \emph{et~al.}
\newblock \bibinfo{title}{{Radial distribution of the multiple stellar
  populations in {$\omega$} Centauri}}.
\newblock \emph{\bibinfo{journal}{\aap}} \textbf{\bibinfo{volume}{507}},
  \bibinfo{pages}{1393--1408} (\bibinfo{year}{2009}).

\bibitem{armitageclarke1996}
\bibinfo{author}{{Armitage}, P.~J.} \& \bibinfo{author}{{Clarke}, C.~J.}
\newblock \bibinfo{title}{{Magnetic braking of T Tauri stars}}.
\newblock \emph{\bibinfo{journal}{\mnras}} \textbf{\bibinfo{volume}{280}},
  \bibinfo{pages}{458--468} (\bibinfo{year}{1996}).

\bibitem{bouvier1997}
\bibinfo{author}{{Bouvier}, J.}, \bibinfo{author}{{Forestini}, M.} \&
  \bibinfo{author}{{Allain}, S.}
\newblock \bibinfo{title}{{The angular momentum evolution of low-mass stars.}}
\newblock \emph{\bibinfo{journal}{\aap}} \textbf{\bibinfo{volume}{326}},
  \bibinfo{pages}{1023--1043} (\bibinfo{year}{1997}).

\bibitem{weiss2000}
\bibinfo{author}{{Weiss}, A.}, \bibinfo{author}{{Denissenkov}, P.~A.} \&
  \bibinfo{author}{{Charbonnel}, C.}
\newblock \bibinfo{title}{{Evolution and surface abundances of red giants
  experiencing deep mixing}}.
\newblock \emph{\bibinfo{journal}{\aap}} \textbf{\bibinfo{volume}{356}},
  \bibinfo{pages}{181--190} (\bibinfo{year}{2000}).

\bibitem{dantonaventura2007}
\bibinfo{author}{{D'Antona}, F.} \& \bibinfo{author}{{Ventura}, P.}
\newblock \bibinfo{title}{{A model for globular cluster extreme anomalies}}.
\newblock \emph{\bibinfo{journal}{\mnras}} \textbf{\bibinfo{volume}{379}},
  \bibinfo{pages}{1431--1441} (\bibinfo{year}{2007}).

\bibitem{piotto2014lp}
\bibinfo{author}{{Piotto}, G.} \emph{et~al.}
\newblock \bibinfo{title}{{The Hubble Space Telescope UV Legacy Survey of
  Galactic Globular Clusters. I. Overview of the Project and Detection of
  Multiple Stellar Populations}}.
\newblock \emph{\bibinfo{journal}{ArXiv e-prints}}  (\bibinfo{year}{2014}).

\bibitem{dicrisci2015}
\bibinfo{author}{{Di Criscienzo}, M.} \emph{et~al.}
\newblock \bibinfo{title}{{An HST/WFC3 view of stellar populations on the
  horizontal branch of NGC 2419}}.
\newblock \emph{\bibinfo{journal}{\mnras}} \textbf{\bibinfo{volume}{446}},
  \bibinfo{pages}{1469--1477} (\bibinfo{year}{2015}).

\bibitem{caloi2007}
\bibinfo{author}{{Caloi}, V.} \& \bibinfo{author}{{D'Antona}, F.}
\newblock \bibinfo{title}{{NGC 6441: another indication of very high helium
  content in globular cluster stars}}.
\newblock \emph{\bibinfo{journal}{\aap}} \textbf{\bibinfo{volume}{463}},
  \bibinfo{pages}{949--955} (\bibinfo{year}{2007}).

\bibitem{bellini63882013}
\bibinfo{author}{{Bellini}, A.} \emph{et~al.}
\newblock \bibinfo{title}{{The Intriguing Stellar Populations in the Globular
  Clusters NGC 6388 and NGC 6441}}.
\newblock \emph{\bibinfo{journal}{\apj}} \textbf{\bibinfo{volume}{765}},
  \bibinfo{pages}{32} (\bibinfo{year}{2013}).

\bibitem{salaris2008}
\bibinfo{author}{{Salaris}, M.}, \bibinfo{author}{{Cassisi}, S.} \&
  \bibinfo{author}{{Pietrinferni}, A.}
\newblock \bibinfo{title}{{The Horizontal Branch of NGC 1851: Constraints on
  the Cluster Subpopulations}}.
\newblock \emph{\bibinfo{journal}{\apjl}} \textbf{\bibinfo{volume}{678}},
  \bibinfo{pages}{L25--L28} (\bibinfo{year}{2008}).

\bibitem{dantona2013}
\bibinfo{author}{{D'Antona}, F.} \emph{et~al.}
\newblock \bibinfo{title}{{The puzzle of metallicity and multiple stellar
  populations in the globular clusters in Fornax}}.
\newblock \emph{\bibinfo{journal}{\mnras}} \textbf{\bibinfo{volume}{434}},
  \bibinfo{pages}{1138--1150} (\bibinfo{year}{2013}).


\vskip 30pt
\bibitem{cardelli1989}
\bibinfo{author}{{Cardelli}, J.~A.}, \bibinfo{author}{{Clayton}, G.~C.} \&
  \bibinfo{author}{{Mathis}, J.~S.}
\newblock \bibinfo{title}{{The relationship between infrared, optical, and
  ultraviolet extinction}}.
\newblock \emph{\bibinfo{journal}{\apj}} \textbf{\bibinfo{volume}{345}},
  \bibinfo{pages}{245--256} (\bibinfo{year}{1989}).

\bibitem{odonnell1994}
\bibinfo{author}{{O'Donnell}, J.~E.}
\newblock \bibinfo{title}{{R$_{nu}$-dependent optical and near-ultraviolet
  extinction}}.
\newblock \emph{\bibinfo{journal}{\apj}} \textbf{\bibinfo{volume}{422}},
  \bibinfo{pages}{158--163} (\bibinfo{year}{1994}).

\bibitem{moehler2011}
\bibinfo{author}{{Moehler}, S.} \emph{et~al.}
\newblock \bibinfo{title}{{The hot horizontal-branch stars in {$\omega$}
  Centauri}}.
\newblock \emph{\bibinfo{journal}{\aap}} \textbf{\bibinfo{volume}{526}},
  \bibinfo{pages}{A136} (\bibinfo{year}{2011}).

\bibitem{ventura2008code}
\bibinfo{author}{{Ventura}, P.}, \bibinfo{author}{{D'Antona}, F.} \&
  \bibinfo{author}{{Mazzitelli}, I.}
\newblock \bibinfo{title}{{The ATON 3.1 stellar evolutionary code. A version
  for asteroseismology}}.
\newblock \emph{\bibinfo{journal}{\apss}} \textbf{\bibinfo{volume}{316}},
  \bibinfo{pages}{93--98} (\bibinfo{year}{2008}).

\bibitem{castcast1993}
\bibinfo{author}{{Castellani}, M.} \& \bibinfo{author}{{Castellani}, V.}
\newblock \bibinfo{title}{{Mass loss in globular cluster red giants - an
  evolutionary investigation}}.
\newblock \emph{\bibinfo{journal}{\apj}} \textbf{\bibinfo{volume}{407}},
  \bibinfo{pages}{649--656} (\bibinfo{year}{1993}).

\bibitem{michaud2007}
\bibinfo{author}{{Michaud}, G.}, \bibinfo{author}{{Richer}, J.} \&
  \bibinfo{author}{{Richard}, O.}
\newblock \bibinfo{title}{{Horizontal Branch Evolution and Atomic Diffusion}}.
\newblock \emph{\bibinfo{journal}{\apj}} \textbf{\bibinfo{volume}{670}},
  \bibinfo{pages}{1178--1187} (\bibinfo{year}{2007}).

\bibitem{michaud2008}
\bibinfo{author}{{Michaud}, G.}, \bibinfo{author}{{Richer}, J.} \&
  \bibinfo{author}{{Richard}, O.}
\newblock \bibinfo{title}{{Abundance Anomalies in Horizontal Branch Stars and
  Atomic Diffusion}}.
\newblock \emph{\bibinfo{journal}{\apj}} \textbf{\bibinfo{volume}{675}},
  \bibinfo{pages}{1223--1232} (\bibinfo{year}{2008}).

\bibitem{gallet2013}
\bibinfo{author}{{Gallet}, F.} \& \bibinfo{author}{{Bouvier}, J.}
\newblock \bibinfo{title}{{Improved angular momentum evolution model for
  solar-like stars}}.
\newblock \emph{\bibinfo{journal}{\aap}} \textbf{\bibinfo{volume}{556}},
  \bibinfo{pages}{A36} (\bibinfo{year}{2013}).

\bibitem{palacios2003}
\bibinfo{author}{{Palacios}, A.}, \bibinfo{author}{{Talon}, S.},
  \bibinfo{author}{{Charbonnel}, C.} \& \bibinfo{author}{{Forestini}, M.}
\newblock \bibinfo{title}{{Rotational mixing in low-mass stars. I Effect of the
  mu-gradients in main sequence and subgiant Pop I stars}}.
\newblock \emph{\bibinfo{journal}{\aap}} \textbf{\bibinfo{volume}{399}},
  \bibinfo{pages}{603--616} (\bibinfo{year}{2003}).

\bibitem{spada2011}
\bibinfo{author}{{Spada}, F.}, \bibinfo{author}{{Lanzafame}, A.~C.},
  \bibinfo{author}{{Lanza}, A.~F.}, \bibinfo{author}{{Messina}, S.} \&
  \bibinfo{author}{{Collier Cameron}, A.}
\newblock \bibinfo{title}{{Modelling the rotational evolution of solar-like
  stars: the rotational coupling time-scale}}.
\newblock \emph{\bibinfo{journal}{\mnras}} \textbf{\bibinfo{volume}{416}},
  \bibinfo{pages}{447--456} (\bibinfo{year}{2011}).

\bibitem{charbonnel2005}
\bibinfo{author}{{Charbonnel}, C.} \& \bibinfo{author}{{Talon}, S.}
\newblock \bibinfo{title}{{Influence of Gravity Waves on the Internal Rotation
  and Li Abundance of Solar-Type Stars}}.
\newblock \emph{\bibinfo{journal}{Science}} \textbf{\bibinfo{volume}{309}},
  \bibinfo{pages}{2189--2191} (\bibinfo{year}{2005}).

\bibitem{castellikurucz2004}
\bibinfo{author}{{Castelli}, F.} \& \bibinfo{author}{{Kurucz}, R.~L.}
\newblock \bibinfo{title}{{New Grids of ATLAS9 Model Atmospheres}}.
\newblock \emph{\bibinfo{journal}{ArXiv Astrophysics e-prints}}
  (\bibinfo{year}{2004}).

\bibitem{dc2008}
\bibinfo{author}{{D'Antona}, F.} \& \bibinfo{author}{{Caloi}, V.}
\newblock \bibinfo{title}{{The fraction of second generation stars in globular
  clusters from the analysis of the horizontal branch}}.
\newblock \emph{\bibinfo{journal}{\mnras}} \textbf{\bibinfo{volume}{390}},
  \bibinfo{pages}{693--705} (\bibinfo{year}{2008}).

\bibitem{ventura2011}
\bibinfo{author}{{Ventura}, P.} \& \bibinfo{author}{{D'Antona}, F.}
\newblock \bibinfo{title}{{Hot bottom burning in the envelope of super
  asymptotic giant branch stars}}.
\newblock \emph{\bibinfo{journal}{\mnras}} \textbf{\bibinfo{volume}{410}},
  \bibinfo{pages}{2760--2766} (\bibinfo{year}{2011}).

\bibitem{cassisi2009wcen}
\bibinfo{author}{{Cassisi}, S.} \emph{et~al.}
\newblock \bibinfo{title}{{Hot Horizontal Branch Stars in {$\omega$} Centauri:
  Clues about their Origin from the Cluster Color Magnitude Diagram}}.
\newblock \emph{\bibinfo{journal}{\apj}} \textbf{\bibinfo{volume}{702}},
  \bibinfo{pages}{1530--1535} (\bibinfo{year}{2009}).

\bibitem{binney1987}
\bibinfo{author}{{Binney}, J.} \& \bibinfo{author}{{Tremaine}, S.}
\newblock \emph{\bibinfo{title}{{Galactic dynamics,}}} 
\newblock \emph{Second Edition} {Princeton University Press (2008)}. 

\end{thebibliography}
\end{document}